\documentclass[preprint,tightenlines,aps,showpacs,letterpaper]{revtex4}

\usepackage{graphicx}
\usepackage{slashed}

\usepackage{amsmath}
\usepackage{amssymb}
\usepackage[hidelinks]{hyperref}

\usepackage{color}

\newcommand{\notE}{ \hbox{{$E$}\kern-.60em\hbox{/}}}
\newcommand{\notp}{\ \hbox{{$p$}\kern-.43em\hbox{/}}}

\preprint{\font\fortssbx=cmssbx10 scaled \magstep2
\hbox to \hsize{
\hskip1.2in 
\hbox{\fortssbx The University of Oklahoma}
\hskip0.2in $\vcenter{
                      \hbox{\bf arXiv: [hep-ph]}
                    \hbox{\bf OU-HEP-201202}
                     \hbox{December 2020}}$ }
}

\begin{document}

\title{\vspace*{0.7in}
Flavor changing top decays to charm and Higgs with $\tau \tau$ at the LHC}

\author{
Phillip Gutierrez\footnote{E-mail address: pgutierrez@ou.edu},
Rishabh Jain\footnote{E-mail address: Rishabh.Jain@ou.edu} and
Chung Kao\footnote{E-mail address: Chung.Kao@ou.edu}
}

\affiliation{
Homer L. Dodge Department of Physics and Astronomy,
University of Oklahoma, Norman, OK 73019, USA }

\date{\today}

\bigskip

\begin{abstract}
  
  We investigate the prospects of discovering the top quark decay into
a charm quark and a Higgs boson ($t \to c h^0$) in top quark pair
production at the CERN Large Hadron Collider (LHC).
A general two Higgs doublet model is adopted to study flavor changing
neutral Higgs (FCNH) interactions.
We perform a parton level analysis as well as Monte Carlo simulations 
using \textsc{Pythia}~8 and \textsc{Delphes} to study the flavor changing
top quark decay 
$t \to c h^0$, followed by the Higgs decaying into $\tau^+ \tau^-$, 
with the other top quark decaying to a bottom quark ($b$) and 
two light jets ($t\to bW\to bjj$). 
To reduce the physics background to the Higgs signal,
only the leptonic decays of tau leptons are considered,
$\tau^+\tau^- \to e^\pm\mu^\mp +\slashed{E}_T$,
where $\slashed{E}_T$ represents the missing transverse energy from
the neutrinos.
In order to reconstruct the Higgs boson and top quark masses as well as 
to reduce the physics background, the collinear approximation
for the highly boosted tau decays is employed.
Furthermore, the energy distribution of the charm quark helps set the
acceptance criteria used to reduce the background and improve the statistical 
significance of the signal. 
We study the discovery potential for the FCNH top decay
at the LHC with collider energy $\sqrt{s} = 13$ and 14 TeV as well as
a future hadron collider with $\sqrt{s} = 27$ TeV.
Our analysis suggests that a high energy LHC at $\sqrt{s} = 27$ TeV 
will be able to discover this FCNH signal with an integrated
luminosity $\mathcal{L} = 3$ ab$^{-1}$ for a branching fraction 
${\cal B}(t \to ch^0) \agt 1.4 \times 10^{-4}$, 
which corresponds to a FCNH coupling $|\lambda_{tch}| \agt 0.023$.
This FCNH coupling is significantly below the current ATLAS combined
upper limit of $|\lambda_{tch}| = 0.064$.

\end{abstract}

\maketitle

\section{Introduction}

The discovery of the Higgs boson in
2012~\cite{Aad:2012tfa,Chatrchyan:2012ufa}
completes the experimental observation of the particle spectrum
predicted by the Standard Model (SM). A primary
goal of the high luminosity and higher energy
Large Hadron Collider (LHC) is the precision testing of the SM
and the search for physics beyond the Standard Model (BSM),
especially the interactions of the Higgs boson, the top quark, and sources
of CP violation. Several experimental
searches~\cite{Sirunyan:2020sum,Aad:2020zxo,Sirunyan:2020xwk,Aad:2020kub}
are being performed to improve the understanding of Higgs boson
interactions with
SM particles and to search for possible extensions of the Higgs
sector.

There are some deviations from the SM such as 
the presence of baryon asymmetry in the
Universe~\cite{Sakharov:1967dj} requiring CP violation beyond that predicted
by the SM.
The muon anomalous magnetic moment measurements
at BNL~\cite{Bennett:2006fi} and Fermilab~\cite{Abi:2021gix}
show approximately 4.2$\sigma$ deviation
from the SM~\cite{Zyla:2020zbs,Borsanyi:2020mff,Aoyama:2020ynm}.
In addition, there might be possible flavor anomalies 
among the quarks and leptons~\cite{
Lees:2013uzd,Huschle:2015rga,Aaij:2015yra,Aaij:2017vbb,Sirunyan:2017dhj}.
The Standard Model with one Higgs doublet cannot explain
these anomalies~\cite{Crivellin:2012ye} thus requiring BSM physics.
A general two Higgs doublet model (2HDM) provides a simple extension
to the SM.
It consists of two scalar SU(2) doublets, which after electroweak symmetry
breaking (EWSB) leads to
five physical Higgs bosons: two CP-even scalars
[$H^0$ (heavier) and $h^0$ (lighter)], one CP-odd pseudoscalar ($A^0$) and
a pair of charged Higgs boson ($H^\pm$).
General 2HDMs can provide additional sources of
CP violation~\cite{Weinberg:1990me,Fuyuto:2017ewj}, and
generate tree-level flavor changing neutral Higgs (FCNH) interactions 
that can enhance the branching fractions of flavor changing neutral
currents, especially $t \to c \phi^0$~\cite{Hou:1991un},
$\phi^0 \to t \bar{c} +\bar{t} c$~\cite{
  Altunkaynak:2015twa,Arroyo-Urena:2019fyd},
and $\phi^0 \to \tau^\pm\mu^\mp$~\cite{
 Khachatryan:2015kon,Aad:2016blu,Sirunyan:2017xzt,Sirunyan:2019shc},
where $\phi^0 = H^0, h^0$ and $A^0$.
The SM expectation is ${\cal B}(t \to c h^0) \approx 10^{-14}$~\cite{
 AguilarSaavedra:2004wm,Mele:1998ag,Eilam:1990zc}, which is significantly
less than current and near term experiments can observe.
If this FCNH signal $t \to c h^0$ is observed at the LHC or HL-LHC,
it would imply BSM physics~\cite{AguilarSaavedra:2000aj,Kao:2011aa,
  Chen:2013qta,Atwood:2013ica,Khachatryan:2014jya,Durieux:2014xla,
  Chen:2015nta,Khachatryan:2016atv,Papaefstathiou:2017xuv,Aaboud:2018oqm,
  Jain:2019ebq,Arroyo-Urena:2019qhl,Castro:2020sba,Zhang:2020naz}.

We adopt the Yukawa Lagrangian in a general
two Higgs doublet model~\cite{Davidson:2005cw,Mahmoudi:2009zx} as 
\begin{equation}\label{eq:yukawaL}
  \begin{aligned}
{\cal L}_Y =& \frac{-1}{\sqrt{2}} \sum_{\scalebox{0.6}{F=U,D,L}}
 \bar{F}\Big\{  \left[ \kappa^Fs_{\beta-\alpha}+\rho^F c_{\beta-\alpha} \right] h^0 +
   \left[ \kappa^Fc_{\beta-\alpha}-\rho^Fs_{\beta-\alpha} \right] H^0 \\
   &- i \, {\rm sgn}(Q_F)\rho^F A^0 \Big\} P_R F 
  -\bar{U} \left[ V \rho^D P_R - \rho^{U\dagger} V P_L \right] D H^+
  -\bar{\nu} \left[ \rho^L P_R \right] L H^+ + {\rm H.c.} \, 
\end{aligned}
\end{equation}
where $P_{L,R} \equiv ( 1\mp \gamma_5 )/2$,
$c_{\beta-\alpha} \equiv \cos(\beta-\alpha)$,
$s_{\beta-\alpha} \equiv \sin(\beta-\alpha)$,
$\alpha$ is the mixing angle between neutral Higgs scalars,
$\tan\beta \equiv v_2/v_1$~\cite{Gunion:1989we} is the ratio of
the vacuum expectation values of the two Higgs doublets,
$Q_F$ is the fermion charge, and the
$\kappa$~matrices are diagonal and fixed by
fermion masses to $\kappa^F = \sqrt{2}m_F/v$ with $v \approx 246$~GeV, while
the matrices $\rho$ contain both diagonal and off-diagonal
elements with free parameters.
In addition, $F,U,D,L$ represent elementary fermions, up-type quarks, down-type
quarks, and charged leptons, respectively. The matrix elements $\rho$
are the FCNH couplings to the fermions.
Almost all experimental data are consistent with the Standard
Model~\cite{Sirunyan:2018koj,Aad:2019mbh},
which implies all two Higgs doublet models must be
in the decoupling~\cite{Gunion:2002zf}
or the alignment limit~\cite{Craig:2013hca,Carena:2013ooa}
with one SM-like light scalar ($h^0$) that has a mass of 125 GeV.

Recently the ATLAS Collaboration~\cite{Aaboud:2018oqm} combined
several channels to search for $t \to c h^0$ with
$h^0 \to b\bar{b}$,
$h^0 \to \tau\tau$ with at least one hadronic tau decay,
$h^0 \to WW^*, ZZ^*, \tau^+\tau^-$ (same sign 2$\ell$, 3$\ell$), 
and $h^0 \to \gamma\gamma$,
and put a strong constraint on the branching fraction
${\cal B}(t \to c h^0) \leq 1.1\times 10^{-3}$. 
This leads to an upper limit on the FCNH Yukawa coupling $|\lambda_{tch}|$
\begin{equation}
    \lambda_{tch} \leq 0.064 \, ,
\end{equation}
for the effective Lagrangian,
\begin{equation}
  \mathcal{L} = -\frac{\lambda_{tch}}{\sqrt{2}}\bar{c}th^0 + H.c.
  \label{eq:effLag}
\end{equation}
with the relation between $\lambda_{tch}$ and
the $t \to c h^0$ branching
fraction~\cite{TheATLAScollaboration:2013nbo} being
\begin{equation}
    \lambda_{tch} \approx 1.92\times\sqrt{\mathcal{B}(t \to c h^0)} \, .
\end{equation}

In this article, we investigate the discovery potential of 
the top quark decay into a charm quark and a Higgs boson ($t \to c h^0$)
followed by the Higgs boson decaying into $\tau^+ \tau^-$ in top quark pair
production at the CERN Large Hadron Collider (LHC).
To investigate the discovery potential of a flavor changing neutral
Higgs boson signal with low physics background, we consider only
the leptonic decays of the tau leptons,
$\tau^+\tau^- \to e^\pm\mu^\mp +\slashed{E}_T$,
where $\slashed{E}_T$ is the missing transverse energy in the event
from the neutrinos.
This is complementary to the ATLAS searches for same charge dileptons.

We perform a parton level analysis as well as a Monte Carlo simulation 
using  \textsc{Pythia}~8~\cite{Sjostrand:2014zea} and
\textsc{Delphes}~\cite{deFavereau:2013fsa} to study the FCNH decay of one
top quark
while the other top quark decays hadronically to a bottom quark ($b$) and 
two light jets: 
$pp \to t\bar{t} \to b W^{\pm} c h^0 \to b j j c \tau^+ \tau^- +X$.
We have calculated the production rates using the full tree level matrix
elements including the Breit-Wigner resonance for both signal and
background process.
In addition, we optimize our acceptance using a standard selection based
technique, as well as using a boosted decision tree to improve the signal
to background ratio and statistical significance.

Since we did not apply charm tagging, our analysis is suitable
for a general search for $t \to q h^0, q = u, c$. Many previous
studies have adopted the Cheng-Sher Ansatz~\cite{Cheng:1987rs}
as the benchmark Yukawa coupling 
\begin{equation}
  \lambda_{tqh} = \frac{\sqrt{2 m_t m_q}}{v}
\end{equation}  
where $q = u, c$ and $v \approx 246$~GeV is the Higgs vacuum
expectation value. The FCNH couplings as the geometric mean for top
and charm quarks is
\begin{equation}
  \lambda_{tch}(CS) = \frac{\sqrt{2 m_t m_c}}{v} \approx 0.0895 \, ,
\end{equation}
which has been excluded by recent ATLAS experiment~\cite{Aaboud:2018oqm}.
For simplicity, we assume $\lambda_{tch} \gg \lambda_{tuh}$ and focus
on the search for $t \to c h^0$. To verify that the associated quark
is a charm, we will need to apply charm tagging.

There are several aspects to note in this analysis.
 To reconstruct the Higgs boson
 and the top quark, the collinear approximation of
 tau decays~\cite{Hagiwara:1989fn} is used.
The collinear approximation for tau decays with physical momentum
 fractions $x_i$ ($0 < x_i < 1$), where $x_i = p(\ell_i)/p(\tau), i = 1, 2$,
 more effectively reduced the physics background than the centrality
 requirement suggested in Refs.~\cite{Chen:2015nta,Aaboud:2018oqm}.
 Furthermore, the energy of the charm quark in the top quark rest frame
 provides good acceptance for the FCNH
 top signal while rejecting background~\cite{Han:2001ap,Kao:2011aa}.
 Promising results are presented for the LHC with $\sqrt{s} = 14$ TeV
 and 27 TeV.

\section{Higgs Signal and Event Selections}

This section presents the cross section for the FCNH signal
$t \to c h^0$ from top quark pair production and outlines our
search strategy for this signal at the LHC.
We focus on the discovery channel with one top quark decaying
hadronically ($t \to b jj$), while the other top quark decays
into a charm quark and a Higgs boson ($h^0$) followed by
$h^0 \to \tau^+ \tau^- \to e^\pm \mu^\mp +\slashed{E}_T$.
Unless explicitly specified,
$q$ generally denotes a quark ($q$) or an anti-quark ($\bar{q}$)
and $\ell^\pm$ will represent an $e^\pm$ or $\mu^\pm$.
This means our FCNH signal has the final state of
$pp \to t\bar{t} \to bjj c e^\pm \mu^\mp +\slashed{E}_T +X$,
where $X$ represents all other particles produced in $pp$ collisions.
Since the mass of the Higgs boson is much greater than the tau lepton's
mass ($M_h \gg m_\tau$), the tau leptons are highly boosted. Therefore,
the collinear approximation of the tau decay~\cite{Hagiwara:1989fn}
is employed to reconstruct the Higgs boson mass and the top quark
mass.

At parton level, our analysis 
employs \textsc{MadGraph}5-aMC-NLO~\cite{Alwall:2011uj}
to calculate tree-level cross sections for the full process
$p p \to t\bar{t} \to bjj \bar{c}h^0 \to bjj \bar{c}\tau^+\tau^- +X$
along with the collinear approximation of tau decays~\cite{Hagiwara:1989fn}. 
The parton level cross section is evaluated using the 
\textsc{CT14LO} parton distribution functions (PDFs)~\cite{Dulat:2015mca}.
For simplicity, the factorization scale ($\mu_F$) and
the renormalization scale ($\mu_R$) are  chosen to be the invariant mass of
the top quark pair ($M_{t\bar{t}}$).
With the above scale choices and PDFs, our
current estimates suggest a $K$-Factor of $\approx 1.8$, and is approximately
the same for all three energies ($\sqrt{s} = 13, 14,\text{and}\ 27$~TeV),
investigated for top quark pair production at the LHC.
The $K$-factors are calculated using TOP++ \cite{Czakon:2011xx}.

  This analysis employees the full tree-level matrix elements
to evaluate the cross section for the FCNH signal and physics background. 
In addition, a consistency check for the tree-level signal cross section
has been performed 
in the narrow width approximation by calculating the cross section
$\sigma(pp \to t\bar{t}
\to t c h^0 \to bjj \, c \ell^\pm_1 \ell^\mp_2 \slashed{E}_T +X)$
as the product of cross section times branching
fractions:
\begin{multline}
\sigma(pp \to t\bar{t} \to bjj \bar{t} + X )
 \times {\cal B}(t\to c h^0) \times {\cal B}(h^0 \to \tau^+ \tau^-)
 \\\times {\cal B}({\tau^+ \to \ell^+_1 \nu_{\ell_1}\bar{\nu_{\tau}}})
 \times {\cal B}({\tau^- \to \ell^-_2 \bar{\nu}_{\ell_2}\nu_{\tau}})\, .
\end{multline}

To evaluate the branching fraction of $t \to c h^0$, 
the effective Lagrangian in Eq.~\ref{eq:effLag} is employed.
The resulting decay width is then obtained as 
\begin{eqnarray}
\Gamma(t \to ch^0)
 =  \frac{|\lambda_{tch}|^2}{32\pi}\times (m_t)\times
      [ (1 + r_c)^2 -r_h^2 ]
      \times \sqrt{1-(r_h + r_c)^2}\sqrt{1-(r_h-r_c)^2} \, ,
\end{eqnarray}
with $r_h = M_h/m_t$ and $r_c = m_c/m_t$.
Assuming that the total decay width of the top quark is
\begin{eqnarray}
\Gamma_t = \Gamma( t\to bW ) +\Gamma( t \to c h^0 ),
\end{eqnarray}
the branching fraction of $t \to c h^0$ is
\begin{equation}
  {\cal B}(t \to c h^0) = \frac{ \Gamma(t\to c h^0) }{ \Gamma_t } \, .
\end{equation}
Comparing this with the Yukawa Lagrangian in Eq.~\ref{eq:yukawaL},
we can express
\begin{equation}
\lambda_{tch} = \tilde{\rho}_{tc}\cos(\beta-\alpha)
\end{equation}
with
\begin{equation}
 \tilde{\rho}_{tc} = \sqrt{\frac{|\rho_{tc}|^2 + |\rho_{ct}|^2}{2}} \, .
\end{equation}
  To present the results, the more convenient free parameters
  $\tilde{\rho}_{tc}$ and $\cos(\beta-\alpha)$ are chosen for the FCNH
  Yukawa couplings.

In the event level analysis, parton level samples are generated
from \textsc{MadGraph} using
\textsc{TauDecay}-UFO~\cite{Hagiwara:2012vz} to model $\tau$ decays,
and then the sample is processed with
\textsc{Pythia}~8~\cite{Sjostrand:2014zea} and
\textsc{Delphes}~\cite{deFavereau:2013fsa}
to generate events with hadronization, showering, and detector effects.
  In addition, the MLM-matching/merging~\cite{Hoche:2006ph} algorithm
  is applied to match the additional hadronized
  jets in each event with partons to avoid double counting jets that
  are generated by
parton showering from final state radiation for all background processes.

To provide a realistic estimate for production rates at the LHC,
we evaluate the cross section for the FCNH signal and physics
background in $pp$ collisions with 
the proper tagging and mistagging efficiencies.
The ATLAS tagging efficiencies~\cite{Aad:2019aic}
are adopted to evaluate the cross section for the FCNH signal and
physics background. 
The $b$ tagging efficiency is 0.7,
the probability that a $c$-jet is mistagged as a $b$-jet ($\epsilon_c$)
is approximately 0.14, while
the probability that a light jet ($u, d, s, g$) is mistagged
as a $b$-jet ($\epsilon_j$) is 0.01.

\subsection{Event Selections}

Our FCNH signal comes from top quark pair production with
one top quark decaying into a charm quark and a Higgs boson
while the other top quark decays to an all hadronic final state.
Every event is required to contain at least four jets,
including exactly one that is identified as a $b$ jet.
In addition, there are two opposite charge
leptons of different flavor ($e^\pm \mu^\mp$) with missing transverse
energy from neutrinos.

We adopt the following basic requirements, which are
similar to the ATLAS and CMS $h^0 \to
\tau^+\tau^-$ studies~\cite{Aaboud:2018pen}.
\begin{itemize}
\item[(i)] Four jets including one $b$-jet with $P_T(b,j) \geq 20$ GeV,
\item[(ii)] $|\eta(b,j)| \leq 2.5$,
\item[(iii)] Two opposite charge leptons with $P_T(\ell) \geq 10$ GeV,
      and $|\eta(\ell)| \leq$2.5,
\item[(iv)] $P_T({\rm leading} \; \ell) \geq 20 $ GeV,
\item[(v)] $\Delta R(\ell \ell,jj,bj,\ell j,\ell b) \geq$ 0.4,
\item[(vi)] $\slashed{E}_T \geq$ 25 GeV,
\item[(vii)] All events containing more than one tagged $b$-jet
      with $P_T \geq  20$ GeV and $|\eta| < 2.5$ are rejected.
\end{itemize}

Since the $b$-quark jet is selected through tagging, this leaves three jets
to be identified as two light-flavor jets and a $c$-quark jet. The two
light-flavor jets, $j_1j_2$, are selected by minimizing $|M_{jj}-m_W|$ and
$|M_{bjj}-m_t|$. The remaining jet is labeled as the $c$-quark jet.
For the event to be correctly reconstructed, $j_1$ and $j_2$ must
result from the decay of a $W$ boson, therefore their invariant
mass distribution
$M_{j_1 j_2}$ peaks at $M_W \approx 80.4$~GeV and
$M_{bj_1 j_2}$ has a peak at $m_t \approx 173.2$ GeV.
Using the ATLAS mass resolution~\cite{Aad:2019mkw},
the reconstructed $W$ and top quark masses are required to lie in the
mass windows 
$\Delta M_{j_1 j_2} = 0.20 M_W$ and
$\Delta M_{b j_1 j_2} = 0.25 m_t$.

\subsection{Higgs Mass Reconstruction}

For the FCNH signal,
$t \to c h^0 \to c \tau^+ \tau^- \to c e^\pm \mu^\mp +\slashed{E}_T$,
the reconstruction is performed two ways:
(a) using the invariant $\tau^+\tau^-$ mass from the Higgs decay and
the invariant mass of $c\tau^+\tau^-$ from the top quark decay, 
which have sharp peaks near $M_H$ and $m_t$, and
(b) as in Ref.~\cite{Jain:2019ebq}, using the cluster transverse
masses of $\ell^+\ell^-$ and $c\ell^+\ell^-$, which have broad
peaks near $M_H$ and $m_t$.

\begin{figure}[htb]
 \centering

 \includegraphics[width=68mm]{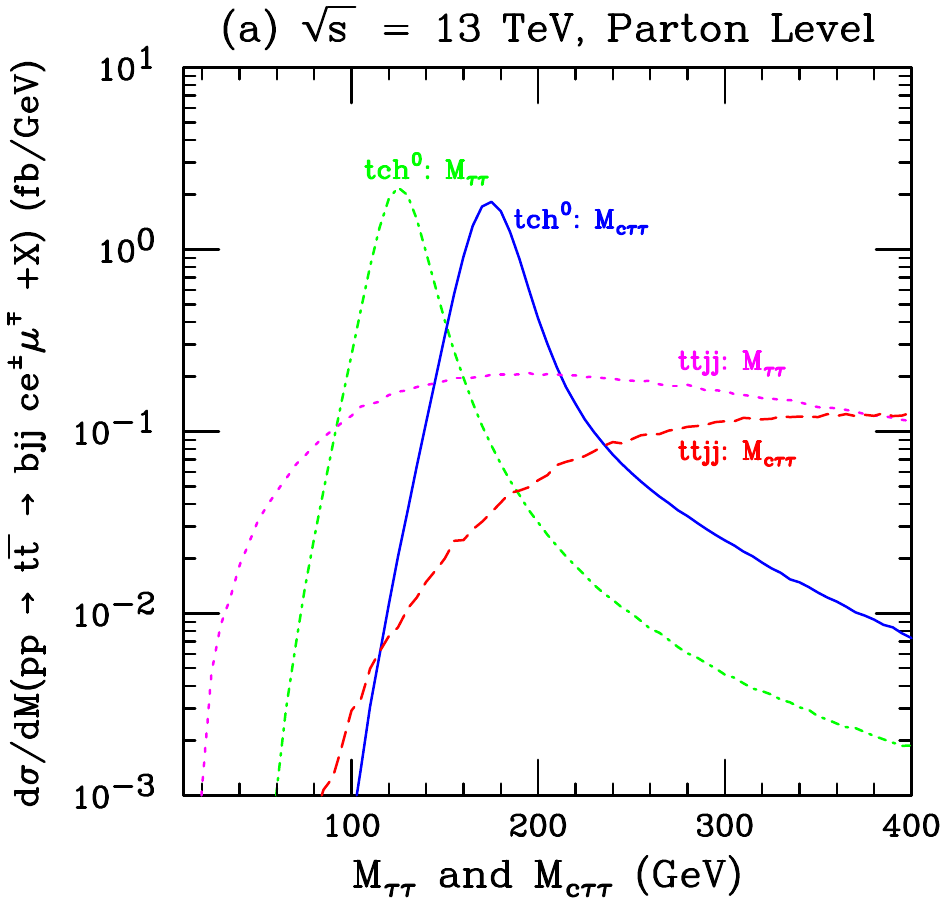}
   \hspace{1mm}
 \includegraphics[width=68mm]{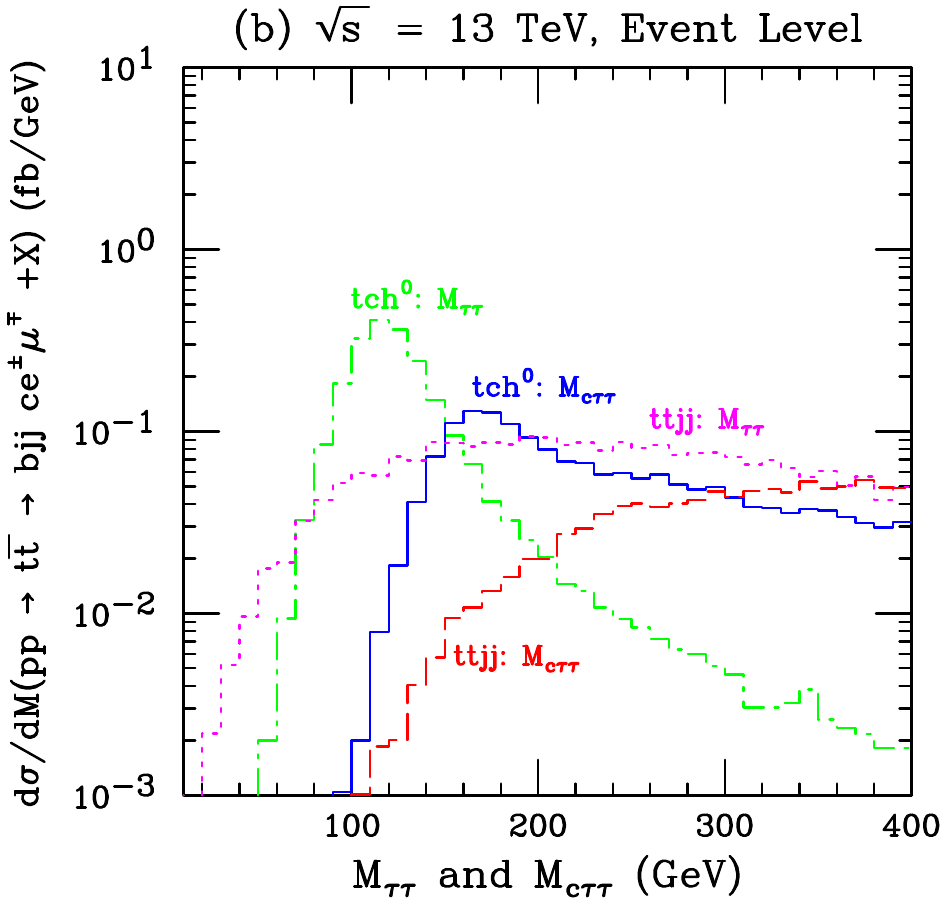}

 \caption{Invariant mass distributions $d\sigma/dM_{\tau\tau}$
   (green dotted dashed)
     and $d\sigma/dM_{c\tau\tau}$ (blue solid)
     for the FCNH signal ($t\to ch^0$) at (a) parton level, and
     (b) event level with detector simulation in $pp$ collisions.
     Also shown are the invariant mass distributions
     $d\sigma/dM_{\tau\tau}$ (magenta dotted)
     and $d\sigma/dM_{c\tau\tau}$ (red dashed) for the dominant
     background from $ttjj$.}
   \label{collinearmass}
\end{figure}

The Higgs boson mass can be reconstructed by applying the collinear
approximation~\cite{Ellis:1987xu,Rainwater:1998kj,Plehn:1999xi}
on the $\tau$ decay products
\begin{equation}
  p_{\ell_i} =  x_i p_{\tau_i} \, , i = 1,2
  \quad {\rm with} \quad p_T(\ell_1) > p_T(\ell_2) \, ,
\end{equation}
and the missing transverse momentum 2-vector, ${\bf \notp_T}$.
Taking $x_i$ to be the momentum fractions carried away from 
the decay by $\ell_i, i = 1, 2$, we have:
\begin{eqnarray}
  \left(\frac{1}{x_1} - 1\right){\bf p_T}(\ell_1)
  +\left(\frac{1}{x_2} - 1\right){\bf p_T}(\ell_2) = 
{\bf \notp_T}.
\end{eqnarray}
This yields two equations for $x_1$ and $x_2$ that can be solved to 
reconstruct the two original $\tau$ 4-momenta 
$p^{\mu}_\tau = p^{\mu}_1/x_1,$ $p^{\mu}_2/x_2$. 
Thus $M_h^2 = (p_1/x_1 + p_2/x_2)^2$ where
$p_1 = p(\ell_1)$ and $p_2 = p(\ell_2)$.
Physically $x_i$ is constrained to $0 < x_i < 1, i = 1,2$, which reduces
the background. 
Figure~\ref{collinearmass} presents the invariant mass distributions 
$M_{col}(\tau \tau)$ and $M_{col}(c,\tau\tau)$,
which have sharp peaks near the Higgs boson and top quark masses,
respectively.
We require
the reconstructed Higgs boson mass and top quark mass to lie in the
mass windows 
$\Delta M_{\tau\tau} = 0.20 M_h$ and
$\Delta M_{c\tau\tau} = 0.25 m_t$ using
the ATLAS mass resolution~\cite{Aaboud:2018pen}.
We note that improvements in the discovery potential are possible 
by reducing the $\tau$ pair mass resolution $\Delta M_{\tau\tau}$.


Furthermore, we employ the cluster transverse mass distributions 
for $e^\pm\mu^\mp$ and $c e^\pm\mu^\mp$  
with missing transverse energy ($\notE_T$) from the neutrinos
to confirm the Higgs boson mass and top quark mass reconstruction.  
These distributions have broad peaks near $M_h$ and $m_t$, respectively,
as the kinematic characteristic of 
$t \to c h^0 \to c\, e^\pm\mu^\mp +\slashed{E}_T$. 
The cluster transverse mass~\cite{Barger:1987nn} is defined as 
\begin{equation}
M_T^2(C,\notE_T)
  =  \left( \sqrt{ p_T^2(C) +M_{C}^2 } + \notE_T \right)^2
      -( \vec{p}_T(C) +\vec{\notE}_T )^2  \, , 
\end{equation}
where $C = \ell^\pm\ell^\mp$ or $c\,\ell^\pm\ell^\mp$,
$p_T(\ell\ell)$ or $p_T(c\,\ell\ell)$ is the total transverse momentum 
of all the visible particles, while $M_{\ell\ell}$ and $M_{c\ell\ell}$ 
are the invariant cluster masses.

In most cases,
the physics background can be reduced and
the statistical significance for the Higgs boson signal enhanced
if we apply a suitable requirement on the cluster transverse mass
distributions~\cite{Jain:2019ebq}
$M_T(\ell\ell,\notE_T)$ and $M_T(c\ell\ell,\notE_T)$.
We have found that acceptance requirement
on $M_{\tau\tau}$ and $M_{c\tau\tau}$ is much more effective
than mass requirement on the cluster transverse masses. 
After the mass selection on the collinear invariant mass,
the effects of additional requirements on the cluster transverse mass
are negligible.

\subsection{Centrality of Missing Transverse Energy}

\begin{figure}
    \centering

    \includegraphics[width=68mm]{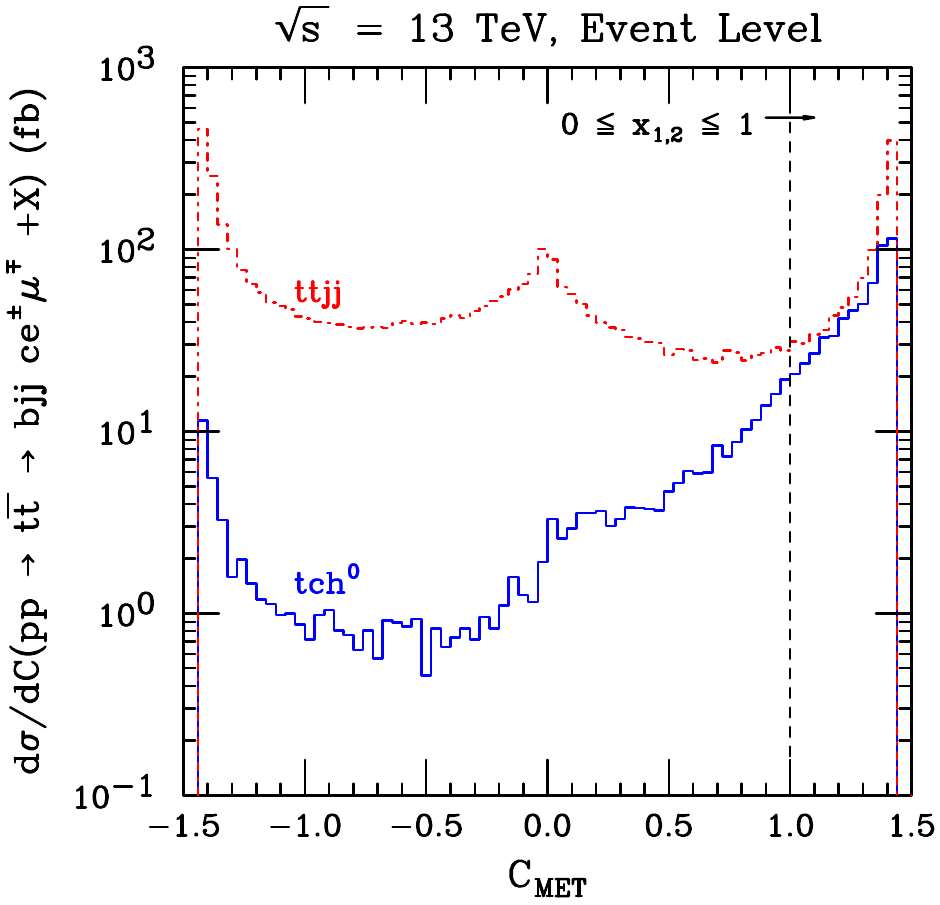}

    \caption{Distribution for the centrality of missing transverse
      energy ($d\sigma/dC_{\rm MET}$)
      for the FCNH signal from
      $t \to c h^0 \to \tau\tau \to e^\pm\mu^\mp +\slashed{E}_T$
      (blue solid) at the LHC with $\sqrt{s} = 13$ TeV. 
      Requiring the momentum fraction to be physical
      $0 \le x_i \le 1, i = 1,2$, we effectively select $C_{\rm MET} >
      1$ that is the region on the right hand slide of the vertical
      dash line.
      Also shown is the centrality distribution of the dominant
      physics background from $ttjj$ (red dotted dashed).}
    \label{fig:centrality}
\end{figure}

To further suppress the physics background, the authors
of Refs.~\cite{Chen:2015nta,Aaboud:2018oqm} suggest the use of
the centrality of the missing transverse energy ($C_{\rm MET}$)
\begin{equation}
C_{\rm MET} = (x+y)/\sqrt{x^2+y^2} \, ,
\end{equation}
with
\begin{equation}
  x = \frac{\sin(\phi_{\rm MET}-\phi_1)}{\sin(\phi_2-\phi_1)} \, ,
  y = \frac{\sin(\phi_2-\phi_{\rm MET})}{\sin(\phi_2-\phi_1)} \, ,
\end{equation}  
where $\phi_{1,2}$ are the azimuthal angles of the two leptons ($e$ or
$\mu$) in the transverse plane, and $\phi_{\rm MET}$ is the azimuthal
angle of the transverse missing energy.
Figure~\ref{fig:centrality} shows the centrality $C_{\rm MET}$ for
the FCNH signal
from $t \to c h^0$ and the dominant background $t\bar{t}jj$.
This is found to be less stringent than the 
requirement on the physical momentum fractions
$0 < x_i < 1, i = 1,2$, which leads to $C_{\rm MET} > 1$.


\section{The Physics Background}

The dominant background to the signal is from $t\bar{t}jj,\ j = q$ or $g$.
Here both top quarks decay leptonically ($t \to b \ell \nu$)
to the desired final state combination of leptons.
This comprises more than 95\% of the total background.
The other dominant contribution is from
$pp \to b\bar{b} jj \tau\tau \to bb jj e^\pm\mu^\mp +\slashed{E}_T +X$
and $p p \to b \bar{b} j j W^+ W^- \to bbjj e^\pm\mu^\mp
+\slashed{E}_T +X$
(without a $t \bar{t}$ contribution) as well as
$t\bar{t} W^\pm $ and $t \bar{t}Z$.
For all of the backgrounds,  one $b$ jet is selected while the other
$b$ jet is mis-identified as a light jet.
Events with two $b$-jets having $p_T(b) > 20$~GeV and
$|\eta(b)| < 2.5$ are vetoed~\cite{Aaboud:2017qyi,Sirunyan:2018lcp}. 
We calculate the
cross section for each of the backgrounds separately using \textsc{MadGraph}
and apply the same event selection procedure as for the signal.
  The irreducible background from
$pp \to t\bar{t}Z +X$ and $pp \to t\bar{t}h^0 +X$ with the subsequent
decay of $Z \to \tau^+\tau^-$ and $h^0 \to \tau^+\tau^-$ are
negligible after all acceptance requirements and the two $b$ veto are
imposed.

We scale our backgrounds to NLO using K-factors of 1.8 for
$t \bar{t} + 2j$, $b \bar{b} j j \tau \tau$, and $b\bar{b}j jW^+W^-$
for all energies i.e. $\sqrt{s}$= 13, 14, and 27 TeV.
For $t\bar{t}W$ and $t\bar{t}Z$  we use the following K-factors
calculated with \textsc{MadGraph}5-aMC-NLO.

\begin{table}[htb]
    \centering
    \begin{tabular}{c|c|c|c}
 Process$\backslash$ $\sqrt{s}$ & 13 TeV & 14 TeV & 27 TeV \\ \hline \hline 
 $t\bar{t}W$ & 1.64 & 1.66 & 1.70 \\
 $t\bar{t}Z$ & 1.46 & 1.49 & 1.50 \\ \hline
      
\end{tabular}
    
    \caption{K-Factors at NLO for $t\bar{t}W$ and $t\bar{t}Z$ produced
    at the LHC.}
    \label{kfactors}
\end{table}

\begin{figure}[htb]
 \centering

 \includegraphics[width=68mm]{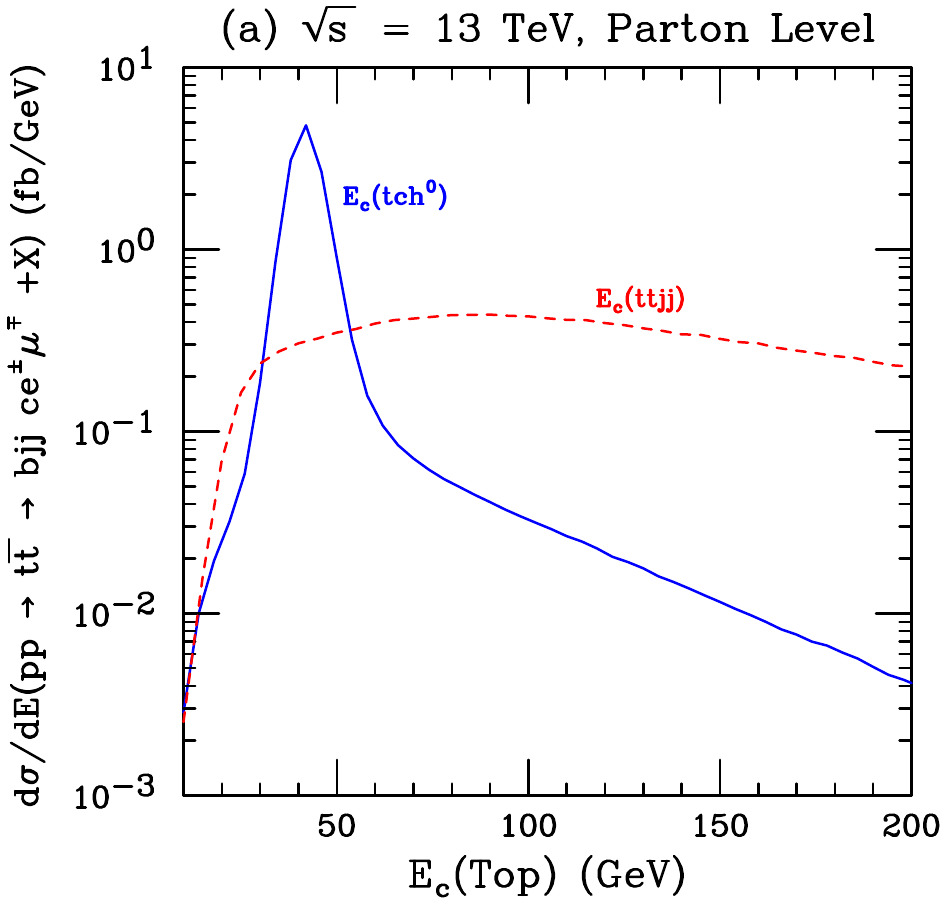}
 \hspace{1mm}
 \includegraphics[width=68mm]{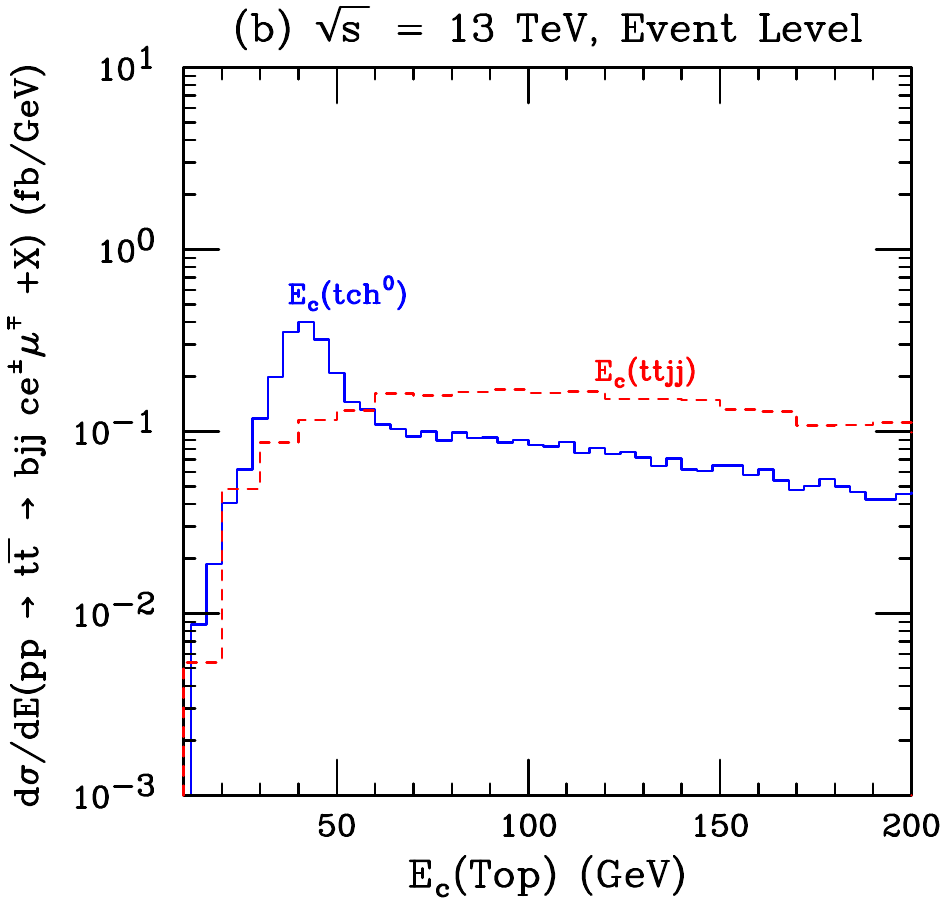}
 
 \caption{Energy distribution for the charm quark ($d\sigma/dE_c$)
   in the top rest frame for the Higgs signal in $pp$ collisions
   with $\sqrt{s} = 13$ TeV, from $t \to c h^0$ (blue solid)
   at (a) parton level, and (b) event level with detector
   simulations. Also shown is the charm quark energy distribution for the
   dominant background $ttjj$ (red dashed).}
 \label{fig:Echarm}
\end{figure}

After applying the event acceptance criteria, we reconstruct the
invariant mass variables $M_{j_1 j_2}$,
$M_{b j_1 j_2}$, $M_{\tau\tau}$, and $M_{c\tau\tau}$,
as discussed in the previous section.
In addition, the energy of the charm quark ($E_c$) in the rest
frame of the top quark is reconstructed to discriminate the  $t \to c h^0$
signal from background~\cite{Han:2001ap,Kao:2011aa}.
For the flavor changing top decay of $t \to c h^0$, the $E_c$
distribution exhibits a peak at the following value,
\begin{equation}
    E_c^* = \frac{m_t}{2}\left[1 + \frac{m_c^2}{m_t^2} 
           -\frac{m_h^2}{m_t^2}  \right] \approx 41.43 \, {\rm GeV} \, .
\end{equation}
Requiring 29 GeV $< E_c <$ 54 GeV, the
background is significantly reduced while most of the signal is maintained.
Figure ~\ref{fig:Echarm} presents the energy distributions of the
charm quark in the top quark rest frame.


From the invariant mass and the charm quark energy distributions  
at the parton and event levels, the following mass requirements
are deduced
\begin{itemize}
\item[(i)] $|M(j_1,j_2) - m_W| \leq 0.20 \times m_W$ and
  $|M(b,j_1,j_2) - m_t | \leq 0.25 \times m_t$,
\item[(ii)] $|M_{col}(\tau,\tau) - m_h| \leq 0.20 \times m_h$ and
  $|M_{col}(c,\tau,\tau) - m_t| \leq 0.25 \times m_t$,
\item[(iii)] 40 GeV$\leq M_{T}(\ell,\ell,\slashed{E}_T) \leq $140 GeV
      and 80 GeV $\leq M_{T}(c,\ell,\ell,\slashed{E}_T) \leq $ 180 GeV, and 
\item[(iv)] 29 GeV $\leq E_c \leq $ 54 GeV.
\end{itemize}
These requirements are chosen to remove the physics background
in a manner that maximizes the statistical significance of the FCNH signal.

\section{Discovery Potential at the Parton Level}

Applying all the selection criteria at parton level for
$\sqrt{s} = 13$, 14 and 27~TeV,
our signal cross sections for $\lambda_{tch} = 0.064$
are shown in Table~\ref{sigcross}.
The cross sections with $\lambda_{tch} = 0.01$ are also presented
for a simple estimate to find the cross sections for other
values of this FCNH Yukawa coupling.
The cross sections for the backgrounds
after applying the selection requirements
are presented in Table~\ref{bkgcross}.

%
%
\begin{table}[htb]
 \centering
 \begin{tabular}{ccc} \hline \hline
 $\sqrt{s}$ (TeV) & $\lambda_{tch} = 0.01$ & $\lambda_{tch} = 0.064$ \\
   \hline \hline  
 13 & 0.0096 & 0.39 \\
 14 & 0.012 & 0.46 \\
 27 & 0.043 & 1.72 \\
 \hline
 \end{tabular}
 \caption{Signal cross section in fb after all cuts,
   scaled with b-tagging = 0.7.}
 \label{sigcross}
\end{table}

%
%
\begin{table}[htb]
  \centering
  \begin{tabular}{cccccccc}
  \hline \hline
  $\sqrt{s}$ (TeV) & $t \bar{t} jj$ & $t \bar{t}jj (+\tau)$ & $b\bar{b}jj\tau\tau$
    & $b\bar{b}j jWW$ & $t \bar{t}jj (+\tau\tau)$ & $t\bar{t}$V & Total  \\
              &                & one $t \to b\tau\nu$ &
    &  & $t\bar{t} \to b\bar{b}\tau^+\tau^-\nu\nu$ &  & \\
    \hline\hline
  13 & 0.45 & 0.21 & 0.021 & 3.2$\times 10^{-4}$  & 0.012 & 3.5$\times 10^{-3}$
     & 0.68  \\
  14 & 0.52 & 0.25 & 0.025 & 3.8$\times 10^{-4}$ & 0.014 & 3.8$\times 10^{-4}$
     & 0.8  \\
  27 & 1.96 & 0.9 & 0.074 & 1.3$\times 10^{-3}$ & 0.05& 9.8$\times 10^{-3}$
     & 2.99  \\ \hline \hline
  \end{tabular}
  \caption{Background cross sections in fb after applying the mass selection
    at the parton level.}
  \label{bkgcross}
\end{table}

\begin{figure}[htb]
    \centering
    \includegraphics[width=80mm]{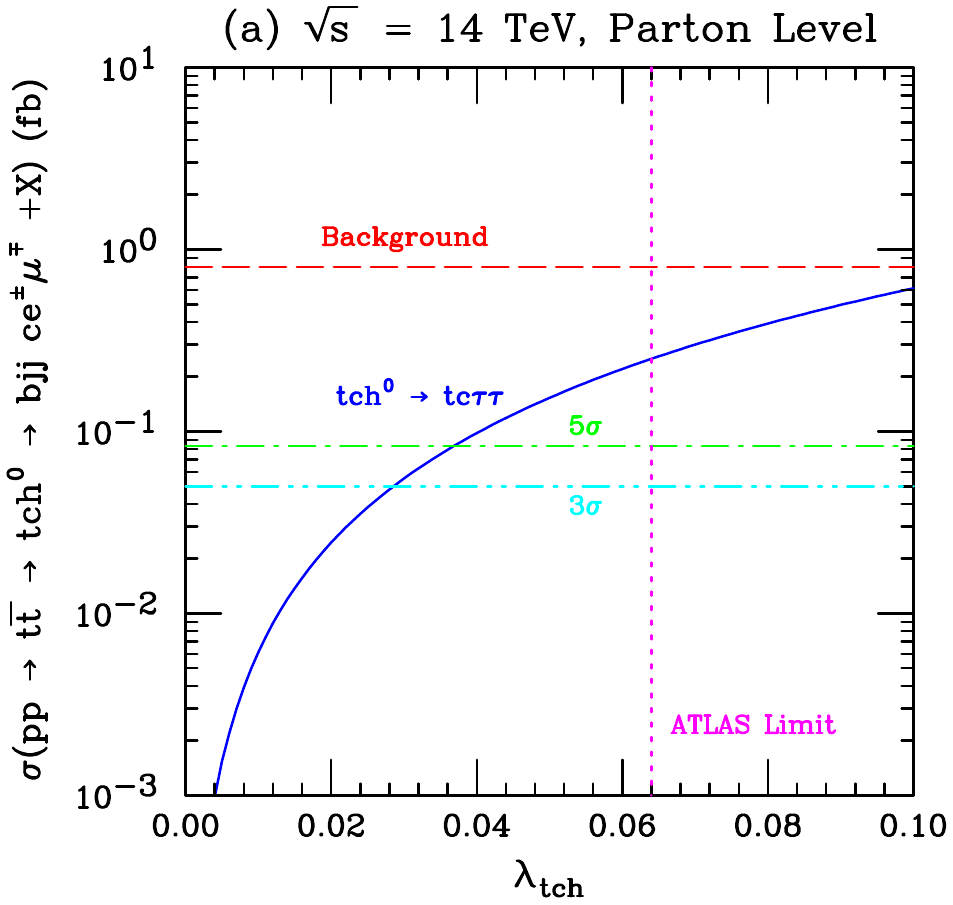}
    \includegraphics[width=80mm]{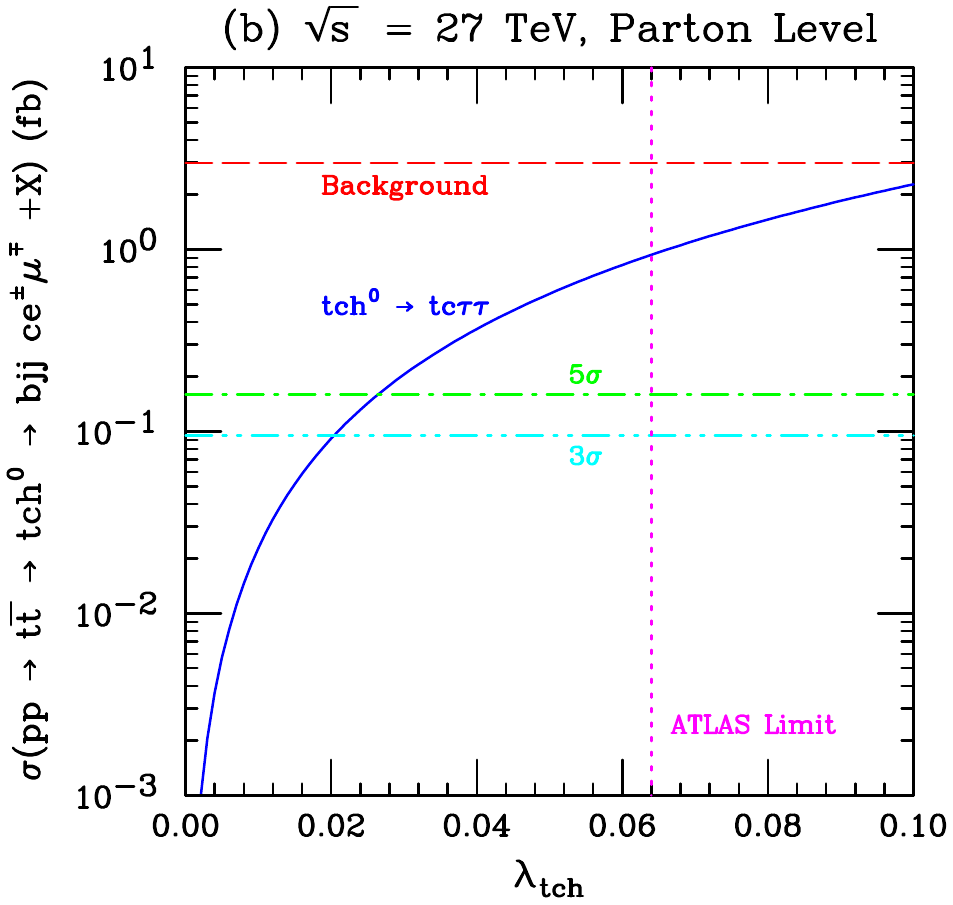} 
    \caption{Cross section of $pp \to tch^0 \to tc\tau\tau \to bjj
      ce^\pm\mu^\mp +\slashed{E}_T +X$ (blue solid) in fb
      as a function of $\lambda_{tch}$
      for $\sqrt{s}=$ (a) 14 TeV and (b) 27 TeV.
      Also shown are the cross section required for $3\sigma$ (cyan
      dotted dotted dashed) and $5\sigma$ (green dotted dashed) as
      well as the cross section of physics background (red dashed).}
    \label{fig:sigVegas}
\end{figure}

Figure~\ref{fig:sigVegas} presents the estimated statistical
significance ($N_{SS}$) as a function of $\lambda_{tch}/\sqrt{2}$ for
the parton level analysis, where
$N_{SS}$ is calculated using \cite{Kumar:2015tna},
\begin{equation}
    N_{SS} = \sqrt{2\times (N_S + N_B)\ln(1 + N_S/N_B) - 2 \times N_S}.
\end{equation}  
Here $N_{S}$ and $N_{B}$ are number of signal and
background events, respectively.


Table~\ref{signcross2} presents a comparison between this study and
our previous study for
$t \to c h^0 \to c W W^*\to c e^\pm\mu^\mp+\slashed{E}_T$~\cite{Jain:2019ebq}.
This analysis
suggests that $h^0 \to \tau \tau$ is much cleaner, because
the Higgs boson mass is fully reconstructed
and the energy of the charm quark in the top quark rest frame
improves the statistical significance using the optimized requirements.

\begin{table}[htb]
    \centering
    \begin{tabular}{ccc} \hline \hline
    $\sqrt{s}$ (TeV) & $h^0 \to WW^*$ & $h^0 \to \tau^+ \tau^-$ \\
      \hline \hline  
    13 & 0.060  & 0.033 \\
    14 & 0.057 & 0.031 \\
    27 & 0.041 & 0.023 \\ 
    \hline
    \end{tabular}
    \caption{Minimum $\lambda_{tch}$ at $\mathcal{L} = 3000$ fb$^{-1}$ for 5 $\sigma$.}
    \label{signcross2}
\end{table}

\begin{figure}[htb]
    \centering
    \includegraphics[width=68mm]{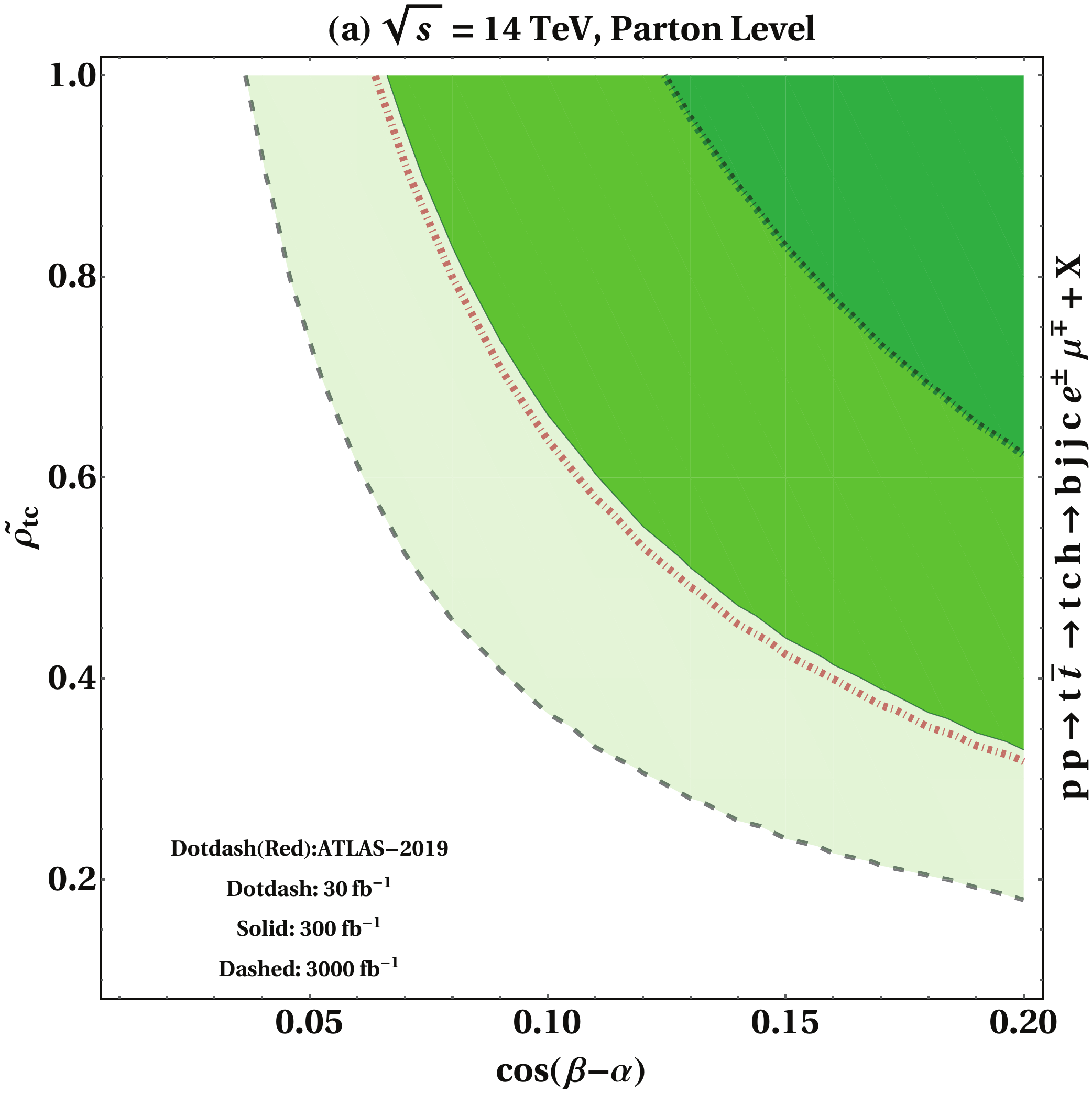}
    \includegraphics[width=68mm]{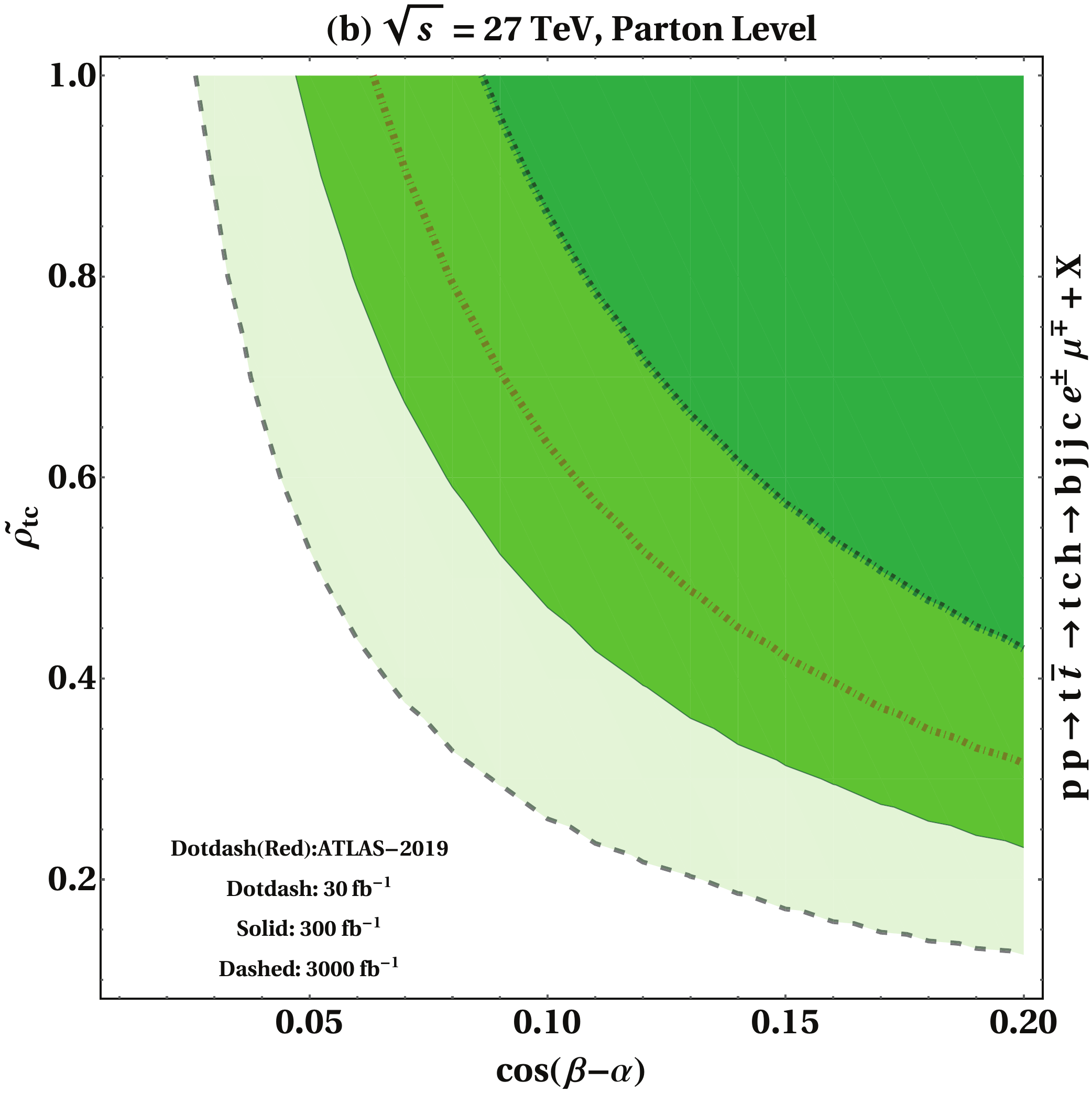} 
    \caption{The parton level 5$\sigma$ discovery contours at the LHC
      in the $[\cos(\beta - \alpha),\tilde{\rho}_{tc}]$ plane
      for (a) $\sqrt{s} =$ 14 TeV and (b) $\sqrt{s} =$ 27 TeV 
      with $L = 30 fb^{-1}$ (dark green dotted dashed),
      $300 fb^{-1}$ (medium green solid) and
      $L = 3000 fb^{-1}$ (light green dashed)
      Also shown is the current limit on
      $\lambda_{tch} = \tilde{\rho}_{tc}\cos(\beta-\alpha)$ (red
      dotted dashed)
      set by ATLAS~\cite{Aaboud:2018oqm}.}
    \label{fig:contours5}
\end{figure}

Figure~\ref{fig:contours5} presents the $5\sigma$ discovery reach at
the LHC for (a) $\sqrt{s} =$ 14 TeV and (b) $\sqrt{s} =$ 27 TeV
at the parton level in the $[\cos(\beta - \alpha),\tilde{\rho}_{tc}]$ plane.
We have chosen $\mathcal{L} =$ 300 and 3000~fb$^{-1}$.
It is clear that the high energy LHC at $\sqrt{s} = 27$ TeV
  with a high luminosity $L = 3000 \; {\rm fb}^{-1}$ significantly
  improves the discovery potential of $t \to ch^0$ for
  $\lambda_{tch} \ge 0.038$ beyond the current ATLAS
  limit~\cite{Aaboud:2018oqm} $\lambda_{tch} = 0.064$.


\section{Event Level Analysis with Boosted Decision Trees}

  In this section,
we present the event level analysis 
using the event generator \textsc{Pythia}~8~\cite{Sjostrand:2014zea} and
the detector simulation program \textsc{Delphes}~\cite{deFavereau:2013fsa}.
From this analysis,
the cross sections for the FCNH signal and the backgrounds
are shown in Table~\ref{crossBSD} after applying the selection
requirements.

%
%
\begin{table}[htb]
    \centering
    \begin{tabular}{cl} \hline \hline
    Process &  Cross-section \\ \hline
    $t\bar{t} jj $    &  1.30 \\
    $b\bar{b}jj \tau\tau$ & 0.07  \\
    $t\bar{t} W$  &  0.008 \\
    $t\bar{t} Z$  &  0.001 \\
    $t\bar{t} h^0$  & 0.0002 \\ 
    $b\bar{b}j j WW$ &  0.001 \\ \hline 
    Total Background & $\approx 1.4$  \\ \hline
    Signal($\lambda_{tch} = 0.01$)  & 0.00098 \\ \hline 
    Signal($\lambda_{tch} = 0.064$) & 0.040 \\ \hline \hline
    \end{tabular}
    \caption{Event level cross section of signal and backgrounds in fb
      at the LHC with $\sqrt{s} = 13$ TeV and all selection requirements.}
    \label{crossBSD}
\end{table}

\begin{figure}
\centering

 \includegraphics[width=68mm]{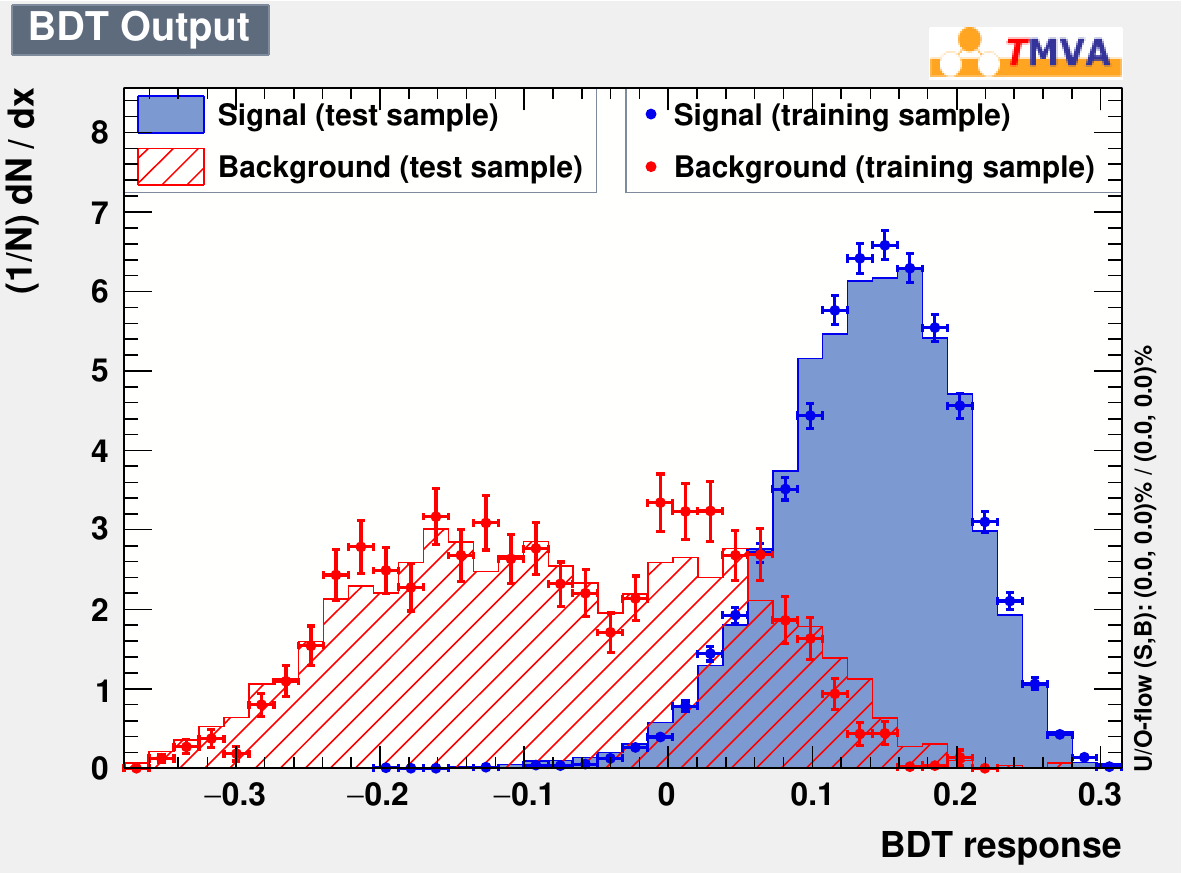}

 \caption{The BDT response for the signal of $t \to c h^0$ and the
   physics background at the event level.}

    \label{bdt}
\end{figure}

For the event level analysis, the mass resolutions are worse than
at the parton level. Therefore, the mass selection window is relaxed
and the selected events are used to
train and test the boosted decision trees (BDT)
classifier to increase the
background rejection relative to signal acceptance.
The \textsc{Root}~\cite{Brun:1997pa} \textsc{TMVA} \cite{Hocker:2007ht}
package is used to perform the signal and background classification.
We apply the following requirements on the sample
\begin{itemize}
 \item[(i)] 65 GeV $\leq M(j_1,j_2) \leq$ 100 GeV,
 \item[(ii)] 40 GeV $ \leq M_T(\ell,\ell) \leq$ 300 GeV,
 \item[(iii)] $M_{col}(\tau,\tau) \leq$ 200 GeV and
       $M_{col}(c,\tau,\tau) \leq $ 300 GeV, and 
 \item[(iv)]  20 GeV $ \leq  E_c  \leq $ 70 GeV,  
\end{itemize}
and then process it through the BDT, which contains 1000 trees
at a depth of 5. The BDT response is shown in Fig~\ref{bdt}.
The BDT is employed to optimize the selection
requirements and improve the statistical significance.

Event selection using the BDT classifier improves the statistical
significance of the analysis relative to using an event based
selection on kinematic and acceptance variables only.
Table~\ref{comp} shows that the BDT analysis improves the statistical
significance by more than a factor of two.

\begin{table}[htb]
    \centering
    \begin{tabular}{cll} \hline \hline
    $\sqrt{s}$ (TeV) & Cut-Based & BDT \\ \hline \hline  
    13 & 1.2 & 2.7 \\
    14 & 1.3 & 3.2 \\
    27 & 2.2 & 5.5  \\
    \hline
    \end{tabular}
    \caption{A comparison of the statistical significance at
      $\lambda_{tch} \approx 0.064$ and $\mathcal{L} = 3000 fb^{-1}$ 
      between a kinematic variable selection analysis (Cut-Based)
      and BDT analysis.}
    \label{comp}
\end{table}

Table~\ref{2sigmaDelphes} presents the 95\% confidence level limits on
$\lambda_{tch}$ at $\sqrt{s} = 13, 14$ and 27 TeV using an
integrated $\mathcal{L} = 300$ and 3000~fb$^{-1}$.
In addition, the
minimum $\lambda_{tch}$ for $5\sigma$ discovery at the LHC is
presented in Table~\ref{5sigdelphes}. We conclude that it will be difficult
to discover this channel at 13 and 14~TeV colliders in this channel,
but a 27 TeV high energy collider holds promise for this signature.

\begin{table}[htb]
    \centering
    \begin{tabular}{cll} \hline \hline
    $\sqrt{s}$ (TeV) & $\mathcal{L} = 300 fb^{-1}$ & $\mathcal{L} = 3000 fb^{-1}$ \\ \hline \hline  
    13 & 0.099 & 0.055 \\
    14 & 0.092 & 0.051 \\
    27 & 0.068 & 0.038 \\
    \hline
    \end{tabular}
    \caption{95 \% C.L Limits on $\lambda_{tch}$ at different
      collider energies and integrated luminosities.} 
    \label{2sigmaDelphes}
\end{table}

\begin{table}[htb]
    \centering
    \begin{tabular}{cll} \hline \hline
    $\sqrt{s}$ (TeV) & $\mathcal{L} = 300 fb^{-1}$ & $\mathcal{L} = 3000 fb^{-1}$ \\ \hline \hline  
    13 & 0.21 & 0.088 \\
    14 & 0.16 & 0.082 \\
    27 & 0.11 & 0.061 \\
    \hline
    \end{tabular}
    \caption{Minimal $\lambda_{tch}$ for $5\sigma$ discovery
      at different collider energies and integrated luminosities.}
    \label{5sigdelphes}
\end{table}

\begin{figure}[htb]
    \centering
    \includegraphics[width=68mm]{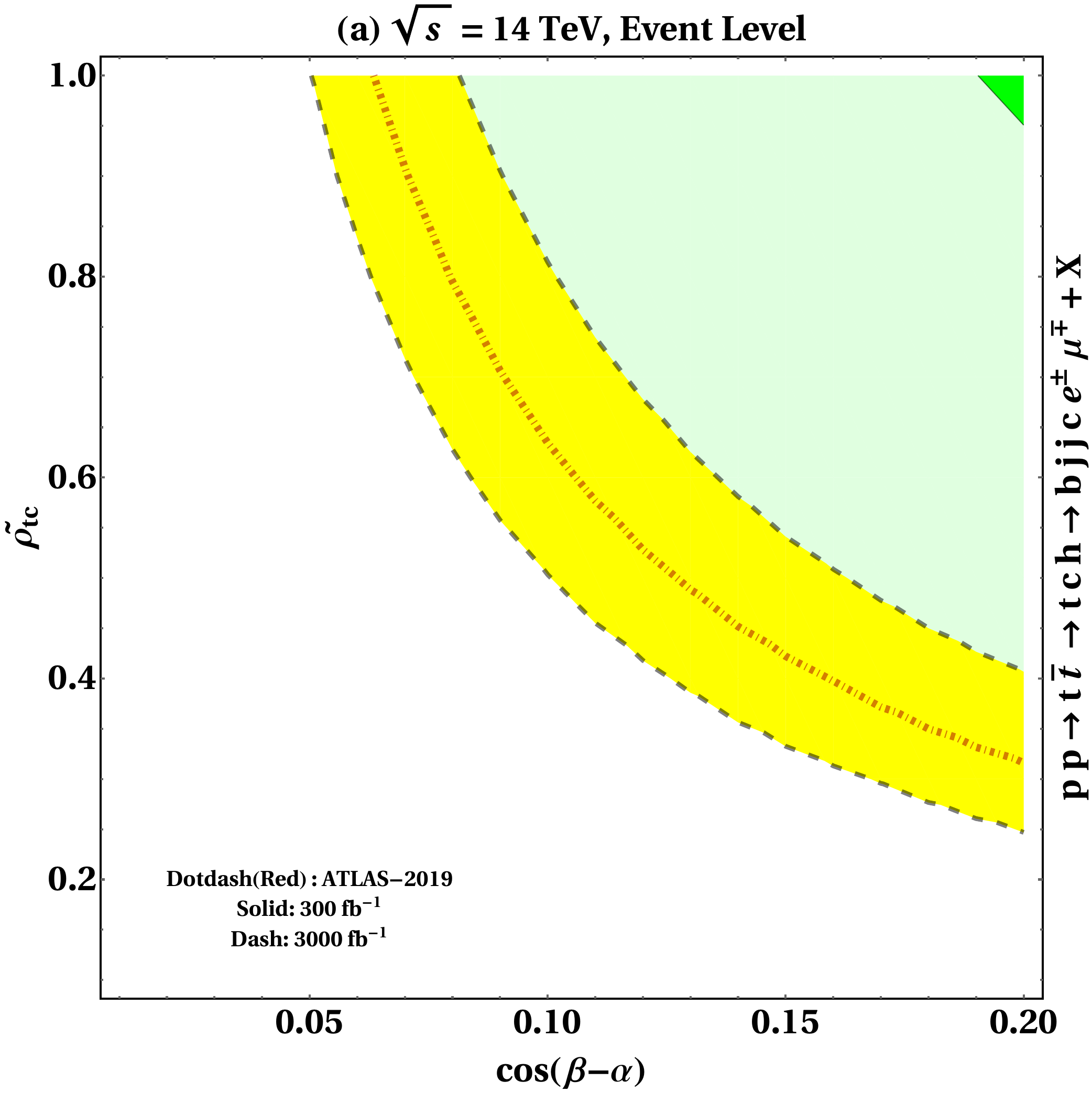}
    \includegraphics[width=68mm]{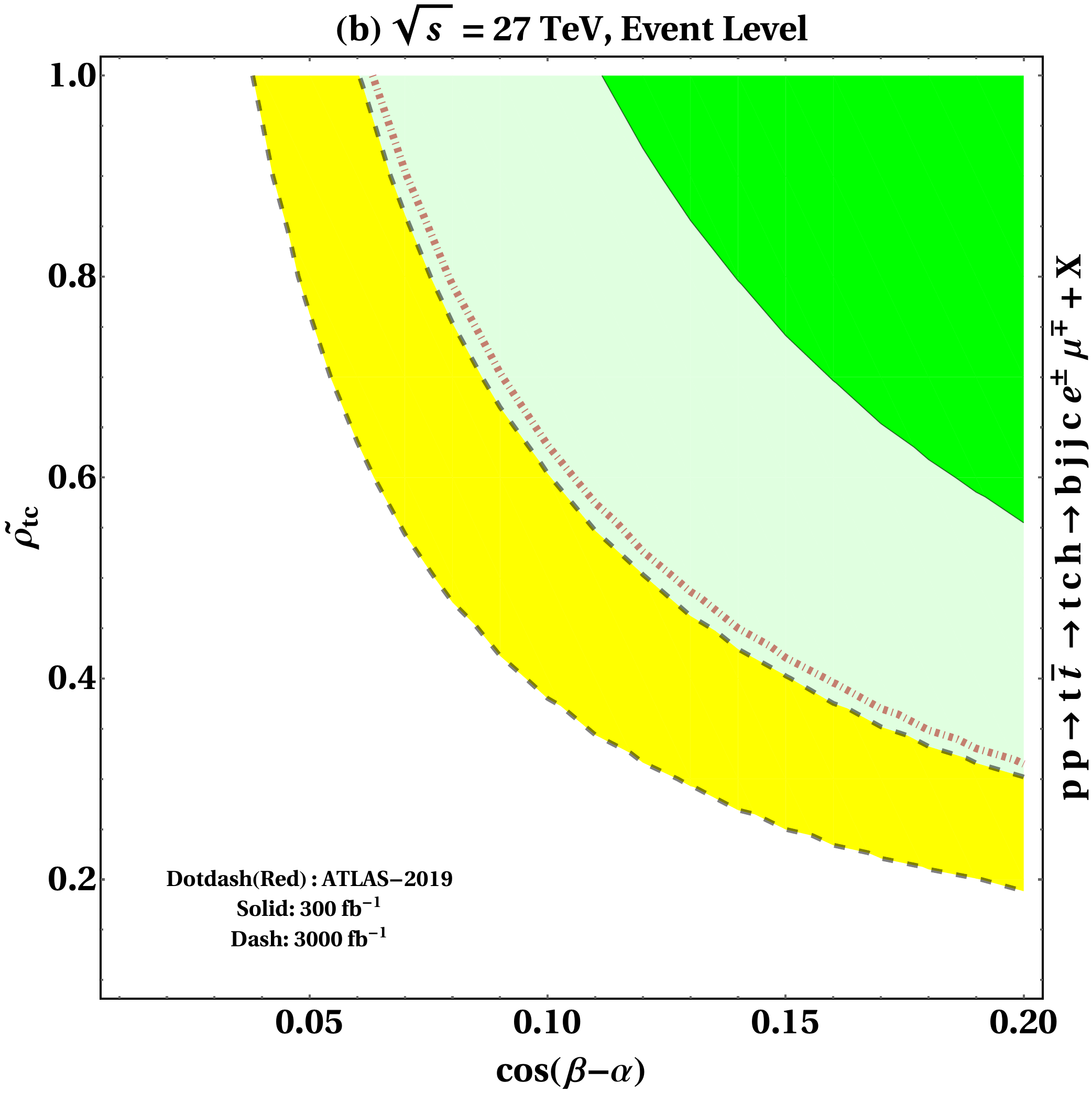}
    \caption{The event level 5$\sigma$ discovery contours at the LHC
      in the $[\cos(\beta - \alpha),\tilde{\rho}_{tc}]$ plane
      for (a) $\sqrt{s} =$ 14 TeV and (b) $\sqrt{s} =$ 27 TeV 
      with $300 fb^{-1}$ (medium green solid) and
      $L = 3000 fb^{-1}$ (light green dashed), as well as the event
      level discovery contours and $3\sigma$ contour (yellow dashed)
      Also shown is the current limit on
      $\lambda_{tch} = \tilde{\rho}_{tc}\cos(\beta-\alpha)$ (red
      dotted dashed)
      set by ATLAS~\cite{Aaboud:2018oqm}.}
    \label{fig:contours7}
\end{figure}

Figure~\ref{fig:contours7} presents the $5\sigma$ discovery reach
at the LHC for
(a) $\sqrt{s} =14$~TeV and (b) $\sqrt{s} =27$~TeV at the event level
in the $[\cos(\beta - \alpha),\tilde{\rho}_{tc}]$ plane.
We have chosen $\mathcal{L} =$ 300 and 3000~fb$^{-1}$.
It is clear that the high energy LHC at $\sqrt{s} = 27$ TeV
  with a high luminosity $L = 3000 \; {\rm fb}^{-1}$ significantly
  improves the discovery potential of $t \to ch^0$
  beyond the current ATLAS
  limit~\cite{Aaboud:2018oqm} $\lambda_{tch} = 0.064$.


  We have illustrated the improvement in statistical significance
  achieved by using a boosted decision trees classifier relative to
  a cut based analysis.
To avoid overtraining the BDT due to low statistics in the event level
analysis, the more restrictive parton level invariant mass
requirements are relaxed.
We then rely on the BDT to optimize the selections on the kinematic variables.
Our goal is to improve the significance by using the BDT to set
the requirements on the invariant masses and the charm-quark energy,
which is a strong signal to background discriminant.
We encourage our experimental colleagues to include the charm-quark
energy as an effective discriminant to further improve the potential of
detecting this FCNH signature at the the LHC.

\section{Conclusions}

Many beyond the Standard Model theories contain tree-level
contributions to FCNH interactions,
especially for the third generation fermions.
These contributions arise naturally in models with additional Higgs
doublets, such as the special two Higgs doublet model for the top quark
(T2HDM)~\cite{Das:1995df},
or a general 2HDM~\cite{Davidson:2005cw,Mahmoudi:2009zx}.
In the decoupling~\cite{Gunion:2002zf} or
the alignment limits~\cite{Craig:2013hca,Carena:2013ooa},
the light Higgs boson ($h^0$) resembles
the standard model Higgs boson with a mass less than the top quark.
This could engender the rare decay $t \to c h^0$.

We investigated the prospects for such a discovery at the LHC,
focusing on the $t\bar{t}$ production channel
and their subsequent decay, where one decays hadronically and the other
through the FCNH mode.
The primary background for this signal is $t\overline{t} j j$
with both top quarks decaying leptonically.
This background contains one $b$ jet mis-identified as a $c$ jet, and
two additional light jets, along with two leptons and missing
transverse energy.
Taking advantage of the available kinematic
information, the $h^0$ and top quark masses in the signal
can be reconstructed and much of the background rejected.

Based on our parton level analysis, we find that the LHC
at $\sqrt{s} = 14$~TeV,
with $\mathcal{L} = 3000$~fb$^{-1}$, can probe to as low as
${\cal B} (t \to c h^0) \approx 2.5 \times 10^{-4}$ with
$\lambda_{tch} \approx 0.033$.
At $\sqrt{s} = 27$~TeV, the reach is ${\cal B}(t \to c h^0)
\approx 1.4 \times 10^{-4}$
with $\lambda_{tch} \approx 0.023$.
The event level analysis implies that there are technical challenges
to reach the discovery potential of the parton level analysis,
especially, improving efficiencies and mass reconstruction with high
precision for final states with missing transverse energy from neutrinos.

In summary, we have made several significant contributions to search
for charming top decays with an associated Higgs boson:
\begin{itemize}
\item[(i)] The $t \to c h^0 \to \tau^+\tau^- \to c e^\pm\mu^\mp$ has not been
  previously investigated as a dedicated discovery channel.
\item[(ii)] We demonstrate the effectiveness of reconstructing the Higgs boson
  and the top quark masses by applying the collinear approximation to the
  tau decays.
\item[(iii)] We show that the requirement on the momentum fractions 
  $0 \le x_i \le 1, i = 1,2$
  is more effective at removing background
  and improving the significance than the requirement on
  centrality ($C_{\rm MET} > 0$).
\item[(iv)] Our requirement on the energy of the charm quark ($E_c$) in the top
  quark rest frame significantly reduces the background
  and improves the significance. 
\item[(v)] We have performed the first investigation of the discovery
  potential of
  $t \to c h^0 \to \tau^+\tau^- \to c e^\pm \mu^\mp$ for a high energy
  $pp$ collider at $\sqrt{s} = 27$ TeV.
\end{itemize}
There are two useful features in the $\tau^+\tau^-$ channel:
(a) the reconstruction of $M_h$ and $m_t$ invariant masses
applying the collinear approximation, and (b) the selection requirement
on the charm quark energy in the top quark rest frame for reducing
the physics background. This leads to the $\tau^+\tau^-$ discovery
channel having a better reach in $\lambda_{tch}$ by a factor of
approximately two over the $W^+ W^-$ channel.

We look forward to being guided by new experimental
results as we explore the interesting physics of EWSB and FCNH.
While the properties of the Higgs boson undergo scrutiny as
data is accumulated, dedicated FCNH searches for $t\to c h^0$ and
$\phi^0 \to t\bar{c} +\bar{t}c, \phi^0 = H^0, A^0$ should be undertaken
for the upcoming high luminosity LHC and future high energy $pp$ colliders.

\section*{Acknowledgments}

C.K. thanks George Hou and the 
High Energy Physics Group at National Taiwan University 
for excellent hospitality, where part of the research was completed.
This research was supported in part by the U.S. Department of Energy.


\end{document}